\documentclass{iau}
\usepackage{graphicx}

\title[Debris Disks in the ALMA era] 
{Debris Disks in Nearby Young Moving Groups in the ALMA Era}

\author[\'A. K\'osp\'al \& A. Mo\'or]   
{\'A. K\'osp\'al$^1$
 \and A. Mo\'or$^1$}

\affiliation{$^1$Konkoly Observatory, Research Centre for Astronomy
  and Earth Sciences, Hungarian Academy of Sciences, PO Box 67, 1525
  Budapest, Hungary \\email: {\tt kospal@konkoly.hu, \tt
    moor@konkoly.hu}}

\pubyear{2015}
\volume{314}  
\pagerange{119--126}
\jname{Young Stars \& Planets Near the Sun}
\editors{J. H. Kastner, B. Stelzer, \& S. A. Metchev, eds.}
\begin{document}

\maketitle

\begin{abstract}
Many members of nearby young moving groups exhibit infrared excess
attributed to circumstellar debris dust, formed via erosion of
planetesimals. With their proximity and well-dated ages, these groups
are excellent laboratories for studying the early evolution of debris
dust and of planetesimal belts. ALMA can spatially resolve the disk
emission, revealing the location and extent of these belts, putting
constraints on planetesimal evolution models, and allowing us to study
planet-disk interactions. While the main trends of dust evolution in
debris disks are well-known, there is almost no information on the
evolution of gas. During the transition from protoplanetary to debris
state, even the origin of gas is dubious. Here we review the exciting
new results ALMA provided by observing young debris disks, and discuss
possible future research directions.  
\end{abstract}

\firstsection

\section{Introduction}
Nearly all young stars are encircled by massive circumstellar disks
that serve as reservoirs for mass accretion and provide the necessary
primordial gas and dust for the formation of planetesimals and
planets. According to observations, gas-rich protoplanetary disks are
depleted within about 10 million years and evolve into gas-poor,
tenuous, dusty debris disks (e.g, \cite[Williams \& Cieza
  2011]{williams2011}). Debris disks are composed of second generation
dust, where individual grains are rapidly removed mainly due to
stellar radiation forces, but the dust is continuously replenished by
collisional erosion or evaporation of previously formed planetesimals
(\cite[Wyatt 2008]{wyatt2008}).

Many young stars ($<$200\,Myr) in the solar neighbourhood belong to
moving groups, i.e., gravitationally unbound, loose associations,
whose members have common origin and move through space together
(e.g., Kastner, this volume). A significant fraction of their members
exhibits excess emission at infrared (IR) wavelengths (e.g.,
\cite[Zuckerman et al.~2011]{zuckerman2011}).  While in some cases
this may come from long-lived protoplanetary disks, the majority can
be attributed to circumstellar debris dust. Many well-studied debris
systems, e.g., $\beta$\,Pic, AU\,Mic, HR\,4796A, belong to young
moving groups. With their proximity and well-dated ages between 8 and
200\,Myr, these groups offer an excellent laboratory for studying the
evolution between the primordial and debris stages, as well as the
early evolution of debris dust and, thereby, of planetesimal
belts. This time range overlaps with the expected last phase of the
formation of terrestrial planets (\cite[Raymond et
  al.~2014]{raymond2014}) and the initiation of collisional cascade
and production of observable debris material in the outer disk regions
(\cite[Kenyon \& Bromley 2008]{kb2008}). However, most debris disks
around members of nearby young moving groups have remained spatially
unresolved so far, providing only limited information about disk
properties.

With its unprecedented sensitivity and spatial resolution, the
recently commissioned Atacama Large Millimeter/submillimeter Array
(ALMA) is opening a new era in the investigation of debris disks by
allowing to observe both the thermal emission of cold dust and the
rotational lines of different gas molecules (if present). By studying
ALMA continuum data we can map the spatial distribution of mm-sized
grains, which dominate the disk emission at submm/mm wavelengths.
These large dust grains are less subject to dynamical removal effects
than smaller grains and are therefore the best tracers of the
structure of the parent planetesimal belts. By determining the
location and extent of these belts, we can put contraints on
planetesimal formation and evolution models and the underlying
dynamical processes. It also allows to probe traces of the interaction
between giant planets and the disk. Molecular line observations with
ALMA allows to measure the amount of gas in debris disks and how it
evolves during the transition from protoplanetary to debris state, a
largely unexplored field so far. In the following we review open
questions concerning the dust and gas content of young debris disks,
the latest ALMA results in the field, and promising future directions.

\section{Stirring of debris disks}

Planetesimals in protoplanetary disks are on low eccentricity and low
inclination orbits because of the damping effect of gas. Therefore,
collisions between them occur at low relative velocities resulting in
the merging of bodies, even after the amount of gas decreases. For
destructive collisions with effective dust production, the motion of
planetesimals needs to be stirred. In the self-stirring scenario
proposed by \cite[Kenyon \& Bromley (2008)]{kb2008}, Pluto-sized
bodies embedded in the outer disk can initiate a collisional cascade
by perturbing the orbits of neighbouring smaller planetesimals. As
large planetesimals build up slower at larger disk radii, the
collisional cascade is ignited in the inner disk first and then the
active dust production propagates outward. As the disk evolves,
planetesimals in the inner regions are ground down, leading to a
decline in the local dust production. This results in an outwardly
increasing dust surface density whose maximum coincides with the
region where Pluto-sized bodies have just been formed (\cite[Kennedy
  \& Wyatt 2010]{kennedy2010}). This is very different from the
surface density profiles observed in protoplanetary disks. The pace of
the outward propagation of the stirring front depends on the disk
surface density: in an initially denser, more massive disk, the
outwards spread is faster (\cite[Kenyon \& Bromley, 2008]{kb2008}). A
giant planet or a stellar companion can also excite the motion of
planetesimals via its secular perturbations (\cite[planetary stirring,
  Wyatt 2005]{wyatt2005}). If the perturber is located closer to the
star than the planetesimal belt, then this mechanism also results in
an inside-out disk stirring whose timescale could be even shorter than
that of self-stirring (\cite[Mustill \& Wyatt 2009]{mustill2009}).

Observational verification of specific aspects of this picture,
however, are not yet conclusive. By mapping the spatial distribution
of mm-sized debris dust grains, we can evaluate several predictions of
stirring models. Using ALMA, \cite[MacGregor et
  al.~(2013)]{macgregor2013} and \cite[Ricci et al.~(2015)]{ricci2015}
successfully resolved the radial density surface density profiles in
two young broad debris disks around AU\,Mic and HD\,107146. In both
cases a rising profile was found at least in the outer regions of the
disks, consistently with the predictions of stirring models. The site
of active dust production in stirring models is expected to propage
outward as a function of age. Previous observations have already
suggested a weak correlation between disk radii and ages (\cite[e.g.,
  Rhee et al.~2007]{rhee2007}; \cite[Eiroa et al.~2013]{eiroa2013}).
However, disk radii in these studies were estimated from their
characteristic temperature and the stellar luminosity, assuming
blackbody grains. Because of the well-known degeneracy between the
equilibrium temperature and size of grains, these estimates are not
unambigous (\cite[e.g. Krivov 2010]{krivov2010}). By measuring debris
disk sizes with ALMA in well-dated systems with different ages, e.g.,
in different young moving groups, this predicted trend can be reliably
verified. In unusually extended young debris disks, the self-stirring
model would require too massive initial disks. In such cases, the
self-stirring scenario is unfeasible, thus, they can be considered as
prime candidates for planetary stirring. Indeed, the debris disks
around HR\,8799, $\beta$\,Pic, and HD\,95086 -- systems with known
outer giant planets -- fall into this category (\cite[Mo\'or et
  al.~2015]{moor2015}). This principle can be used to search for
additional planetary stirring candidates with ALMA. Since the
potential targets are young, using advanced adaptive optics systems
such as the Gemini Planet Imager or VLT/SPHERE (Macintosh et al.~2008,
Beuzit et al.~2010), even the perturber(s) could be identified.

\section{Signatures of planet-disk interaction}

Debris dust grains and the planetesimals from which they are derived
are the smallest constituents of a planetary system that may include
more massive planets as well. As the system evolves, the planet(s) and
the planetesimal disk can interact in many ways. The different types
of gravitational perturbations can cause various kind of footprints,
such as offsets, spirals, gaps, or clumps in the disk. Secular
perturbations from a misaligned or eccentric planet can cause warps or
tightly wound spirals in the disk (\cite[Matthews et
  al.~2014a]{matthews2014a}). These structures then propagate through
the disk, as the gravitational effect of the planet extends to outer
and outer regions. By forcing the distant planetesimals on
intersecting orbits, this may result in more frequent collisions that
occur with higher relative velocities, eventually leading to the
initiation of a collisional cascade. After a time, the secular
perturbations from an eccentric planet make the whole disk also to be
eccentric (e.g.,\cite[Kalas et al.~2005]{kalas2005}).

Resonant perturbations occur when dynamical frequencies, typically of
the mean motions, are a simple integer ratio of each other. Resonances
can lead to either stabilization or destabilization of the orbits,
causing overpopulated or underpopulated regions in a debris disk
(\cite[Matthews et al.~2014a, and references
  therein]{matthews2014a}). As a consequence of overlapping
first-order mean-motion resonances, the region close to the planet's
orbit becomes chaotic, and dust particles orbiting in this zone are
short-lived. Therefore, the chaotic zone is evacuated and a gap is
formed. In our Solar System, the Kirkwood gaps in the asteroid belt
are formed via such mean motion resonances with Jupiter. Overpopulated
regions can also be formed via resonant captures, such as the Trojan
asteroids (in 1:1 resonance with Jupiter) or the Plutinos (in 3:2
resonance with Neptune) in our Solar System. In a debris disk, two
different migration mechanisms are proposed to lead to resonant clumps
in the dust distribution. Outward migration of a planet can trap
planetesimals into its external resonances, the erosion of these
planetesimals then produce an enhanced dust population. Smaller grains
can escape from these regions because of radiation pressure. Large
particles, however, are trapped in the resonance and cause a local
enhancement that can be traced in submillimeter/mm maps of the disk
(\cite[Wyatt, 2003]{wyatt2003}). Alternatively, in those tenuous
debris disks where the timescale of Poynting-Robertson drag is shorter
than the collisional timescale, dust grains produced in an outer
planetesimal belt can spiral inward due to drag forces into the
mean-motion resonances with an inner planet (\cite[Kuchner \& Holman,
  2003]{kh2003}).

ALMA observations of Fomalhaut's disk showed that the mm-sized dust is
concentrated in a very narrow, vertically thin ring, whose inner and
outer edges are sharply truncated (\cite[Boley et
  al.~2012]{boley2012}). This ring morphology can be best explained
with the presence of two shepherding planets located at the inner and
the outer edges, which are likely to be less massive than
3\,M$_\oplus$. HD\,107146 is encircled by a very broad debris disk.
ALMA images of the radial distribution of dust revealed a decrease in
the surface brightness profile at intermediate radii (\cite[Ricci et
  al.~2015]{ricci2015}). One possible interpretation is that this is
related to a gap carved by a few Earth-mass planet in the dusty disk.

Previously, based on spatially resolved millimeter images of nearby
bright debris disks around Vega, $\epsilon$\,Eri, and HD\,107146,
several clumpy structures were reported and the presence of these
inhomogeneities were linked to large grains trapped in mean motion
resonances with unseen planets (\cite[Holland et
  al.~1998]{holland1998}; \cite[Greaves et al.~1998]{greaves1998};
\cite[Corder et al.~2009]{corder2009}). Most of these structures,
however, turned out to be dubious, because subsequent more sensitive
and higher resolution observations could not confirm their presence
(e.g., \cite[Pi\'etu et al.~2011]{pietu2011}; \cite[Hughes et
  al.~2011]{hughes2011}). Thanks to its excellent sensitivity, ALMA
can detect possible resonant structures of even lower contrast in the
brightness distribution of debris disks. Recent ALMA maps of
$\beta$\,Pic revealed a bright region in the southwestern part of the
disk. This clump be seen both in the distribution of CO gas and
mm-sized dust grains (\cite[Dent et al.~2014]{dent2014}) and it also
coincides well with an enhancement of micron-sized grains observed at
resolved mid-IR images (\cite[Telesco et al.~2005]{telesco2005}). Both
the gas and dust in $\beta$\,Pic may have secondary origin. Based on
current data, the clump can be attributed to frequently colliding icy
planetesimals trapped in a mean motion resonance with an outwardly
migrating planet. Alternatively, the clump may be a result of a recent
collision between two Mars-sized icy comets (\cite[Dent et
  al.~2014]{dent2014}).
 
The observed disk asymmetries and structures can be used to deduce the
orbit and other parameters of the planet(s) responsible for them and
can also provide an opportunity to constrain the dynamical history of
the planetary system. In young systems, wide separation giant planets
predicted in this way can be confirmed using new generation adaptive
optics systems. Moreover, such observations of the perturbers will be
able to help in the evaluation and improvement of methods we use in
the interpretation of structures seen in the dust distribution. Indeed
now we may already know two potential benchmark systems. HR\,8799 and
HD\,95086 are two young A-type stars with outer giant planets detected
by direct imaging (\cite[Marois et al.~2010]{marois2010}; \cite[Rameau
  et al.~2013]{rameau2013}). Both systems harbor a warm inner dust
belt and a broad cold outer disk. The giant planet(s) are located
between the two dusty regions (\cite[Matthews et
  al.~2014b]{matthews2014b}; \cite[Mo\'or et al.~2013a]{moor2013};
\cite[Su et al.~2009, 2015]{su2015}). ALMA observations of these disks
are ongoing.

\section{Gas in young debris disks}

Debris disks are expected to be gas-poor systems with a gas-to-dust
ratio significantly lower than in protoplanetary disks. The detection
of the presumably little amounts of gas is a challenging task. Despite
substantial efforts, we know only eight debris disks where gas has
been detected. In the pre-ALMA era, the edge-on orientation of disks
around $\beta$\,Pic and HD\,32297 allowed the detection of their gas
content via absorption lines (\cite[Slettebak et
  al.~1975]{slettebak1975}; \cite[Hobbs et al.~1985]{hobbs1985};
\cite[Redfield 2007]{redfield2007}). The gaseous disks around
HD\,172555, HD\,181296, and AU\,Mic were identified based on O\,I ,
C\,II, and fluorescent H$_2$ line emissions, respectively
(\cite[Riviere-Marichalar et al.~2012, 2014]{rm2012}; \cite[France et
  al.~2007]{france2007}), while CO gas has been observed in disks
around 49\,Ceti, HD\,21997, and HD\,131835 (\cite[Zuckerman et
  al.~1995]{zuckerman1995}; \cite[Mo\'or et al.~2011]{moor2011};
Mo\'or et al.~in prep). All of these systems are probably younger than
50\,Myr, and with the exception of HD\,32297 they belong to young
moving groups or associations. Interestingly, apart from the M-type
AU\,Mic, all of the host stars are A-type. The origin of gas in these
disks is uncertain. In most cases the gas might be secondary, i.e.,
similarly to dust grains it is continuously replenished via erosion of
larger bodies. Collisions between icy planetesimals, evaporation of
comets or icy grains, and photon-induced desorption of solids can lead
to gas release. However, because of their youth, we cannot exclude
that some of these disks are hybrid in the sense that they retain
their residual primordial gas, while the dust component have secondary
origin.

The first observations of gaseous debris disks with ALMA provided
examples of both types. By mapping the dust and gas in the disk around
$\beta$\,Pic, \cite[Dent et al.~(2014)]{dent2014} found that the
spatial distribution of the two components is similar, both showing a
characteristic bright clump in the southwest part of the disk. The
short lifetime of CO and other dynamical arguments make it probable
that the gas in this system has secondary origin, probably indicating
intense collisional activity (\cite[Fern\'andez et
  al.~2006]{fernandez2006}; \cite[Dent et al.~2014]{dent2014}). ALMA
observations of HD\,21997 showed that the CO gas is not co-located
with the mm-sized dust and thereby the planetesimals, but there is a
dust-depleted inner gas disk (\cite[K\'osp\'al et
  al.~2013]{kospal2013}; \cite[Mo\'or et
  al.~2013b]{moor2013b}). Moreover, the disk's total CO content is
more than a thousand times larger than that of $\beta$\,Pic. These
cannot be explained with secondary gas models, implying that this disk
instead harbors residual primordial gas and is the first example of a
hybrid disk (\cite[K\'osp\'al et al.~2013]{kospal2013}). HD\,21997 may
be a unique object or may be the tip of an iceberg of many fainter
hybrid disks. In the latter case, the gas-rich phase of disk evolution
could be significantly longer than previously thought, also affecting
the formation of planets.

\section{Conclusions}

Before ALMA, less than a dozen debris disks were spatially well
resolved at millimeter wavelengths. These studies constituted the
first steps towards understanding the evolution of young debris
systems. ALMA is now in the process of multiplying the sample of
resolved debris disks, thereby allowing to study more general
trends. The first observations have already produced remarkable
results. ALMA confirmed that the rising brightness profile seen in
certain debris disks is consistent with the stirring models. It also
identified disks where the derived initial mass was too high for the
self-stirring scenario, pointing to planetary stirring. It imaged
structures that hint at so-far unknown planets shepherding the dust or
clearing gaps. Finally, it detected a large amount of molecular gas,
leading to the discovery of the first hybrid disk where the dust is
secondary but the gas component is still primordial. In the coming
years, using an increasingly larger sample, we will be able to answer
several fundamental questions concerning the dust and gas content of
debris disks. Dust continuum observations will be used to, e.g.,
verify whether the location of the dust-producing region expands with
time, to identify asymmetric dust distributions and their relationship
with planets. Gas line observations with ALMA are the best way to
study the cold gas content of debris disks, providing a possibility to
study how the gas evolves, and what mechanism produces the gaseous
material. These exciting new results ahead of us will transform the
research of debris disks, and -- in connection with the age and
environmental information provided by their membership in young moving
groups -- will shed light onto a period of very active planet
formation.

\section*{Acknowledgment}
\noindent This work was supported by the Momentum grant of the MTA CSFK
Lend\"ulet Disk Research Group. A.M.~acknowledges support from the
Bolyai Research Fellowship of the Hungarian Academy of Sciences.

\end{document}